# Selective Switch for Band-Edge Alignments in Pentagonal CdSeTe Monolayer: Atom-Valley Locking

Zhi-Qiang Wen [1], Qiu Yang [1], Shu-Hao Cao [1], Zhao-Yi Zeng [2,*], Hua-Yun Geng [3], Xiang-Rong Chen [1]

[1] Institute of Atomic and Molecular Physics, Sichuan University, Chengdu 610065, China;

[2] College of Physics and Electronic Engineering, Chongqing Normal University, Chongqing 401331, China;

[3] National Key Laboratory for Shock Wave and Detonation Physics Research, Institute of Fluid Physics, CAEP, Mianyang 621900, China

## Abstract:

We propose a novel concept named atom-valley locking (AVL) base on two-dimensional (2D) non-magnetic pentagonal $CdSe_2$ monolayer. Here, "locking" refers to an atom-dependent valley distribution. Protected by the four-fold rotational reflection symmetry (S4) solely, two sets of valley electrons can be distinguished both in real space and momentum space. Next, we construct a ternary CdSeTe monolayer to introduce an intrinsic electric field (EF) into the structure without destroying the AVL feature. In momentum space, a spontaneous polarization of 187 meV does not involve time-reversal symmetry can be observed. In real space, each set of the valley electrons can be associated with a distinct band-edge alignment (BEA). Then, we design a selective switch for BEAs via strain engineering, offering precise control of the valley-selective and face-specific hydrogen evolution reaction (HER) site. These results suggest that valleytronics hold great potentials in the field of photocatalytic water splitting.

[*] Corresponding author. E-mail: zhaoyizeng@cqnu.edu.cn

**Keywords**: Atom-valley locking; Valleytronics; Photocatalytic water splitting; Band-edge alignment; Valley polarization

Valleytronics is an emerging field that utilizes the valley degree of freedom (d. o. f.) of electrons to meet specific application needs [1–6]. Recently, based on novel valley behaviors reported in two-dimensional (2D) magnetic materials, a lot of valleytronic devices for different purposes have been designed [7–12]. For example, a spontaneous valley polarization observed in ferrovalley materials [13–18] can be used for non-volatile memory devices. The spin-valley locking observed in altermagnets [19–23] enabling the generation of spin-polarized currents via the valley Hall effect. However, 2D magnetic materials faced a lot of challenges. Firstly, suitable candidates are extremely rare due to the Curie temperature requirement. Secondly, the impact of complex external magnetic fields limits their operating environment. Thirdly, instability and demagnetization in air creates problems for their application. Therefore, exploring novel valley behaviors in non-magnetic systems is beneficial.

Form solar energy to hydrogen production, photocatalytic water splitting [24] is considered a promising solution to the deteriorating energy crisis. This process involves the transition of photogenerated electrons to the conduction band and the creation of holes in the valence band [25]. Most of these carriers accumulate at the valley state and then participate in the hydrogen evolution reaction (HER) and oxygen evolution reaction (OER), with the site determined by valley distribution. Although several novel valley phenomena that couples the valley d. o. f. with spatial d. o. f. have been reported before [26–34]. The great potential of valleytronics devices in photocatalytic applications remains unexplored.

In this work, we introduce a novel concept named atom-valley locking (AVL) and predict ideal candidates via first-principles calculations, specifically the pentagonal $CdSe_2$ monolayer [35] and its family. Here, "locking" refers to an atom-dependent distribution of valley electrons, which is exclusively driven by four-fold rotational reflection symmetry (S4) and does not require intrinsic magnetism. As illustrated in Fig. 1(a), two sets of valley electrons can be distinguished both in real space and momentum space. Next, we construct a ternary pentagonal CdSeTe monolayer to introduce an intrinsic electric filed (EF) into the system while maintaining the AVL feature. As shown in Fig. 1(b), the breaking of S4 symmetry results in a spontaneous valley polarization, which does not involve time-reversal symmetry. Meanwhile, as shown in Fig. 1(c), due to the out-of-plane asymmetry, each set of the valley electrons corresponds to a distinct BEA. Based on these above, we design a selective switch for BEAs via strain engineering, which enables precise control of the valley-selective and face-specific HER site at the atomic scale, as shown in Fig. 1(d). These findings establish a new paradigm for applying valleytronics in the field of photocatalytic water splitting.

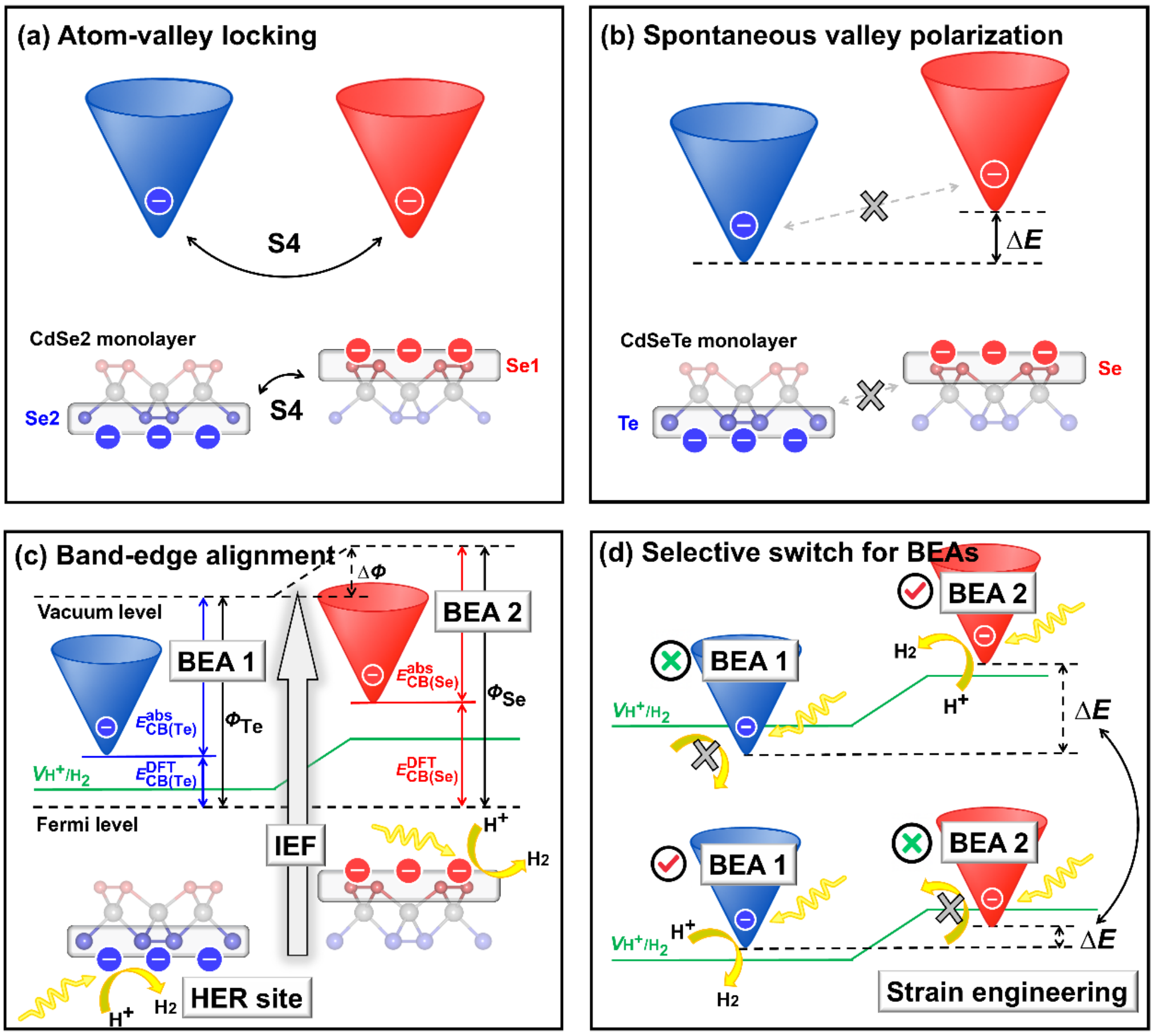

**FIG. 1.** (Color online) (a) The mechanism of atom-valley locking in the $CdSe_2$ monolayer. (b) A spontaneous valley polarization does not involve time-reversal symmetry in the CdSeTe monolayer. (c) Two band-edge alignments induced by intrinsic electric field, each of them corresponds to a distinct valley. (d) A selective switch for band-edge alignments through strain engineering, providing precise control of the valley-selected and face specific hydrogen evolution reaction site.

To establish the locking between atoms in real space and valleys in momentum space in a non-magnetic system, a structural symmetry must serve as a bridge between the two dimensions. We present such symmetry requirements directly: (1) In real space, inequivalent sublattices must be exclusively connected by S4 symmetry. (2) In momentum space, the high-symmetry paths that host the valleys must also be exclusively protected by S4 symmetry.

Through symmetry analysis, we find that the 2D pentagonal structures satisfy requirement (1). Using first-principles calculations, we then screened a total of 72 candidates that meet requirement (2). Their geometric parameters and electronic band structures at the PBE level are presented in section II in the Supplemental Material [36]. The lattice constants of these structures range from 3.3 to 7.5 Å, with thickness ranging from 1.08 to 3.82 Å. Some are semiconductors, with bandgaps ranging from 0.02 to 4.88 eV, while others are semimetals or metals. Some structures satisfy requirement (2) in the conduction band, some in the valence band, and some in both. Although they differ from one another, all of them, without exception, exhibit an atom-dependent valley distribution. These results validate the concept of AVL we proposed and confirm that it is a phenomenon purely driven by structural symmetry. Further investigation beyond the pentagonal configuration will be carried out in future work.

Next, we take the $CdSe_2$ monolayer as an example to discuss the realization of AVL in detail. The optimized crystal structure and structural symmetry are shown in Fig. 2(a), and the obtained structural parameters, including lattice constant, thickness, and bond length are summarized in Table I. In real space, it has a square unit cell with three sublayers, where the Cd sublayer is sandwiched between two Se sublayers, forming a sandwich structure. It belongs to the $P\bar{4}2_1\mathrm{m}$ (No. 113) space group, which includes two perpendicular mirror symmetry planes and an S4 axis passing through the Cd atoms. The Se atoms in the upper sublayer are equivalent with respect to the mirror plane (110), forming the Se1 sublattice. The Se atoms in the lower sublayer are equivalent with respect to the mirror plane ($\bar{1}10$), forming the Se2 sublattice. The two sublattices are inequivalent and can only be connected through the S4 symmetry, thus satisfying requirement (1).

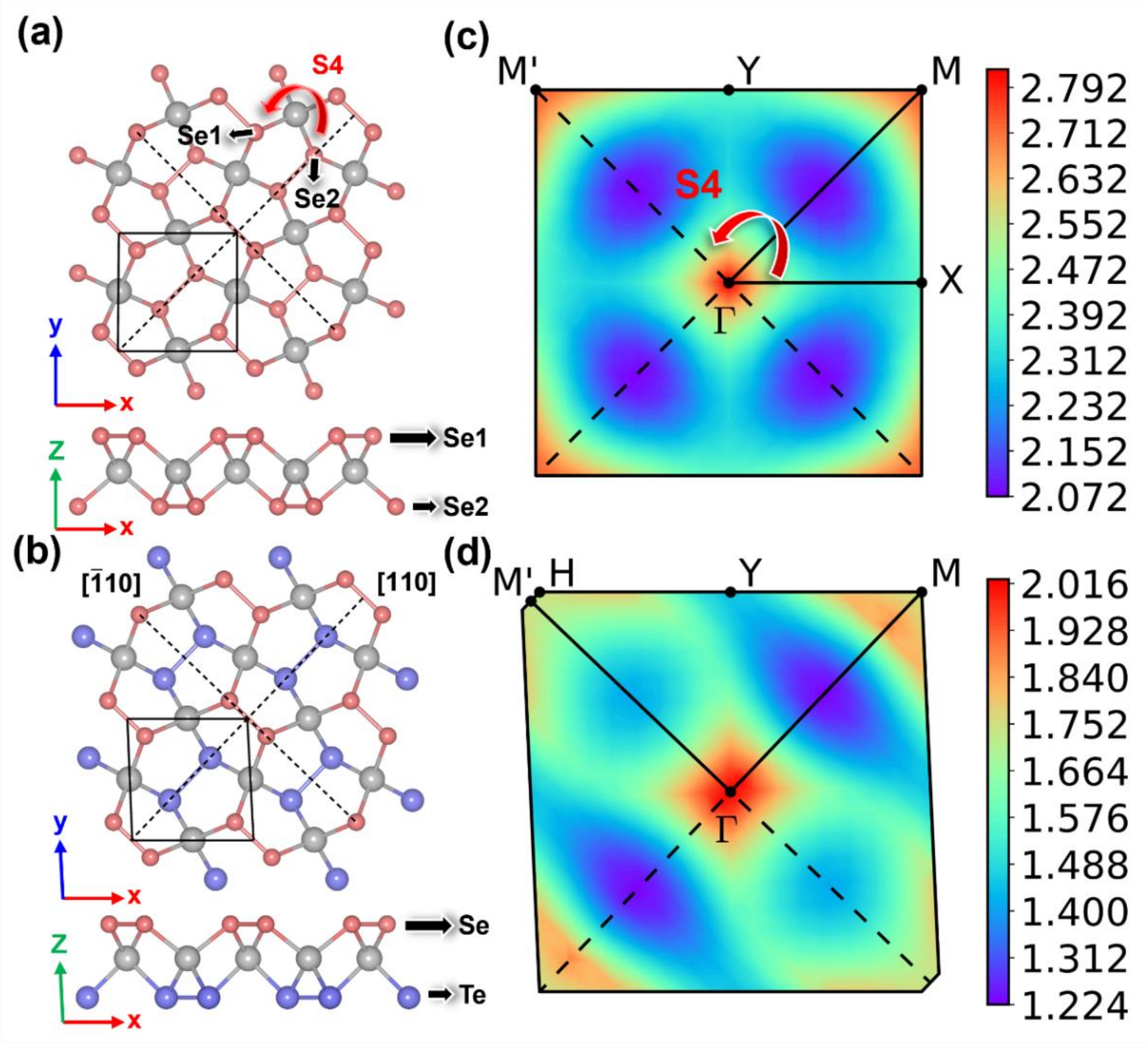

**FIG. 2.** (Color online) The optimized crystal structures of (a) $CdSe_2$ and (b) CdSeTe monolayers. The black solid lines represent the unit-cell, the black dashed lines indicate two perpendicular mirror symmetry planes. The red arrow marks the four-fold rotational reflection axis through Cd atom. The heatmap of the conduction band edge of (c) $CdSe_2$ and (d) CdSeTe monolayers throughout the first Brillouin zone. The solid black lines represent the irreducible Brillouin zone, while the dashed black lines and red arrow indicate their structural symmetry.

**TABLE I.** The obtained lattice constants $a$ (Å), thicknesses $h$ (Å) and bond lengths ($d_{\text{Cd-Se}}$, $d_{\text{Cd–Te}}$, $d_{\text{Se–Se}}$, and $d_{\text{Te–Te}}$) (Å) of the $CdSe_2$ and CdSeTe monolayers.

| Materials | $a$ | $h$ | $d_{\text{Cd–Se}}$ | $d_{\text{Cd–Te}}$ | $d_{\text{Se–Se}}$ | $d_{\text{Te–Te}}$ |
|---|---|---|---|---|---|---|
| $CdSe_2$ | 5.54 | 3.31 | 2.67 | – | 2.40 | – |
| CdSeTe | 5.72 | 3.46 | 2.69 | 2.83 | 2.40 | 2.77 |

The heatmap of the conduction band edge of the $CdSe_2$ monolayer throughout the first Brillouin zone (BZ) is shown in Fig. 2(c). In momentum space, under the constraints of the structural symmetry, the square first BZ can be reduced to a triangular irreducible BZ (Γ–$X$–$M$–Γ) [65]. The high-symmetry path Γ–$X$ is protected by mirror symmetry, while the high-symmetry path $X$–$M$ is protected by both mirror and S4 symmetry. The special high-symmetry path Γ–$M$, which lies on the mirror plane, is protected only by the S4 symmetry. The calculation results show that the valleys occur precisely along this path, thus satisfying requirement (2).

Therefore, the atoms in real space and the valleys in momentum space are directly connected by the S4 symmetry in the $CdSe_2$ monolayer. The atom-resolved projected band structure and the partial charge density of the valleys at the HSE06 level are shown in Figs. 3(a) and 3(c). It can be seen that the material is a semiconductor with an indirect band gap of 2.27 eV. The two valleys are energetically degenerate under the protection of S4 symmetry, but their valley physics are completely different: one is locked by Se1, and the other is locked by Se2. The two sets of valley electrons are distinguishable both in real space and in momentum space.

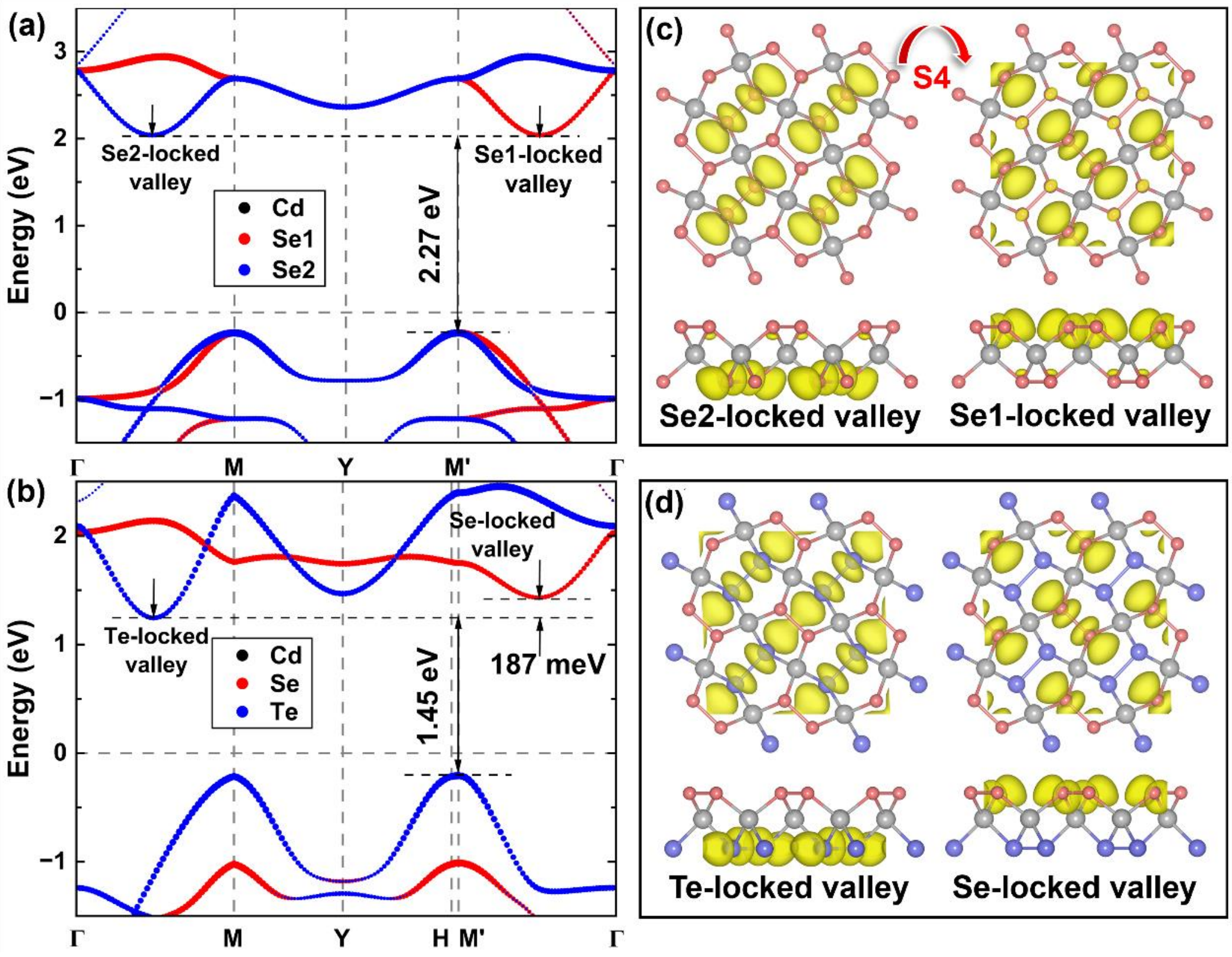

**FIG. 3.** (Color online) The atom-resolved projected band structures of (a) $CdSe_2$ and (b) CdSeTe monolayers at the HSE06 level. The partial charge density of the valleys in the (c) $CdSe_2$ and (d) CdSeTe monolayers. The isosurface value is set to 0.004 $e/Å^3$. The red arrow indicates the S4 symmetry.

In order to introduce an intrinsic EF into the structure without destroying the AVL feature, we design a ternary CdSeTe monolayer with out-of-plane asymmetry by replacing Se2 with Te in the binary $CdSe_2$ monolayer. The calculation results indicate that the target monolayer is dynamically stable, thermodynamically stable, mechanically stable, and thermodynamically stable. It lies at a low position on the equipotential surface and holds promise for synthesis via a bottom-up approach. The corresponding details are presented in section III in the Supplemental Material [36].

The crystal structure and structural symmetry of the CdSeTe monolayer are shown in Fig. 2(b), and the obtained structural parameters are listed in Table I. In real space, it has a rhombic unit cell, with the upper and lower sublayers composed of Se and Te atoms, respectively. It belongs to the *Cmm*2 (No. 35) space group. Compared to its binary counterpart, the mirror symmetry is preserved, while the S4 symmetry is broken. As a result, Se atoms are equivalent to one another, just as Te atoms are, with each forming its own sublattice. However, no symmetry remains that can connect these two sublattices together.

The heatmap of the conduction band edge of the CdSeTe monolayer throughout the first BZ is shown in Fig. 2(d). In momentum space, under the constraint of mirror symmetry, the hexagonal first BZ can be reduced to a trapezoidal irreducible BZ ($\Gamma$–$M$–$Y$–$H$–$M'$–$\Gamma$) [66]. In this case, the high-symmetry paths $\Gamma$–$M$ and $\Gamma$–$M'$ are unprotected by any symmetry. When valleys appear along these two paths, a spontaneous valley polarization of 187 meV is induced. Most importantly, the time-reversal symmetry remains intact due to the absence of intrinsic magnetism or external magnetic field. Giving the fact that magnetic materials faced many challenges in device fabrication, this phenomenon holds significant potential for applications where a stable polarized valley state is required [67,68].

The energy of the valley state is highly sensitive to symmetry breaking. However, previous studies have shown that valley-related locking phenomena are robust under certain perturbations [69,70]. The CdSeTe monolayer can be considered a perturbation of its binary counterpart due to their similar geometric and electronic configurations. The atom-resolved projected band structure and the partial charge density of the valleys at the HSE06 level are shown in Figs. 3(b) and 3(d). The results indicate that the material is a semiconductor with an indirect band gap of 1.45 eV. Two energetically polarized valleys can be found in the conduction band, with one locked by Se and the other locked by Te, thus preserving the AVL feature. Using the same substitution strategy, it is possible to obtain hundreds of ternary pentagonal monolayers based on 72 candidates we provide, which simultaneously exhibit atom-dependent valley distribution and an intrinsic EF. Additionally, the ground-state electronic band structure obtained by different methods including $GW$ and spin-orbit coupling (SOC) can be found in section IV in the Supplemental Material [36] for comparison.

Now, an intrinsic EF is introduced in the AVL-featured CdSeTe monolayer. These two key factors from the foundation of its unique advantages in terms of BEA control. The BEA refers to the requirement that the energy of valley states must overcome to straddle the redox potentials of water [71–75], allowing the valley electrons to participate in the photocatalytic water splitting reaction. The BEA can be described by the following equations [76–78]:

$$E_{\mathrm{CB}}^{\mathrm{abs}} = E_{\mathrm{CB}}^{\mathrm{DFT}} - \Phi > V_{\mathrm{H^+/H_2}} = -4.44 + pH \times 0.059 \text{ eV}, \quad (1)$$

$$E_{\mathrm{VB}}^{\mathrm{abs}} = E_{\mathrm{VB}}^{\mathrm{DFT}} - \Phi < V_{\mathrm{O_2/H_2O}} = -5.67 + pH \times 0.059 \text{ eV}, \quad (2)$$

$$E_{\mathrm{CB}}^{\mathrm{abs}} - E_{\mathrm{VB}}^{\mathrm{abs}} = E_{\mathrm{CB}}^{\mathrm{DFT}} - E_{\mathrm{VB}}^{\mathrm{DFT}} + \Delta\Phi > V_{\mathrm{H^+/H_2}} - V_{\mathrm{O_2/H_2O}} = 1.23 \text{ eV}, \quad (3)$$

where $E_{\mathrm{CB/VB}}^{\mathrm{abs}}$ is the absolute energy of valley states in the conduction band or valence band related to the vacuum level, $E_{\mathrm{CB/VB}}^{\mathrm{DFT}}$ is the energy of valley states in the conduction band or valence band relative to the Fermi level from DFT calculations, $\Phi$ is the work function, which is the difference between the vacuum level and the Fermi level, $\Delta\Phi$ is the difference of the work functions between two surfaces of the material, $V_{H^+/H_2}$ and $V_{O_2/H_2O}$ is the redox potentials of water.

The role of AVL is to provide two sets of valley electrons, which are distinguishable both in real space and momentum space, thus presenting valley-selected and face-specific HER site in the CdSeTe monolayer, as shown in Fig. 1(c). To participate in the photocatalytic reaction, electrons form the Se-locked valley only need to overcome the work function of the upper surface $\Phi_{Se}$, while electrons from the Te-locked valley only need to overcome the work function of the lower surface $\Phi_{Te}$. The role of the intrinsic EF is to provide unevenly distributed work functions between the upper and lower surfaces ($\Phi_{Se} \neq \Phi_{Te}, \Delta\Phi = \Phi_{se} - \Phi_{Te} = 0.26$ eV), as shown in section V in the Supplemental Material [36]. Additionally, there is also an energy polarization between the valley electrons ($E_{CB(Se)}^{DFT} \neq E_{CB(Te)}^{DFT}, \Delta E = E_{CB(Se)}^{DFT} - E_{CB(Te)}^{DFT} = 187$ meV). Therefore, two distinct BEAs exist in the CdSeTe monolayer, each of them is associated with a different valley, and when their energy $E_{CB}^{DFT}$ satisfies equation (1), the corresponding BEA is activated. For structures where AVL appears in the valence band, Equation (2) should be considered, and for structures where AVL appears in both the conduction and valence bands, both Equation (1) and Equation (2) should be considered. Since the performance of the valence band in the CdSeTe monolayer is trivial, the discussion below will focus solely on the conduction band and the HER site to maintain a clear physical picture.

For the first time, direct control of the BEAs by manipulation of valley polarization is achievable. We apply in-plane uniaxial strain along the [110] and [$\bar{1}$10] directions (ranging from -8% to 8%) on the CdSeTe monolayer. It should be noted that we applied such high level of strain is to explore the theoretical possibility of reversing the valley polarization, and is not required for the application scenario discussed later in this work. As shown in Fig.S11 (section VI in the Supplemental Material [36]), the dynamic stability is confirmed for all strain conditions (except for -8% along the $[\bar{1}10]$ direction). Note that the application of compressive strain also induces out-of-plane buckling at the macroscopic scale, and the corresponding molecular dynamics simulations are also provided in section VI in the Supplemental Material [36]. The bandgaps and the valley polarizations $\Delta E$ under all strain conditions at the HSE06 level are presented in Figs. 4(a) and 4(b). As mentioned earlier, CdSeTe monolayer can be considered a perturbation of its binary counterpart. When tensile strain is applied along the [110] direction or compressive strain along the $[\bar{1}10]$ direction, the unit cell of the structure changes from a parallelogram to a square. As the perturbation decreases, valley polarization also diminishes, and two critical conditions (8% along [110] and -6% along $[\bar{1}10]$) can be observed where $\Delta E$ becomes zero, which occurs within the dynamically stable range.

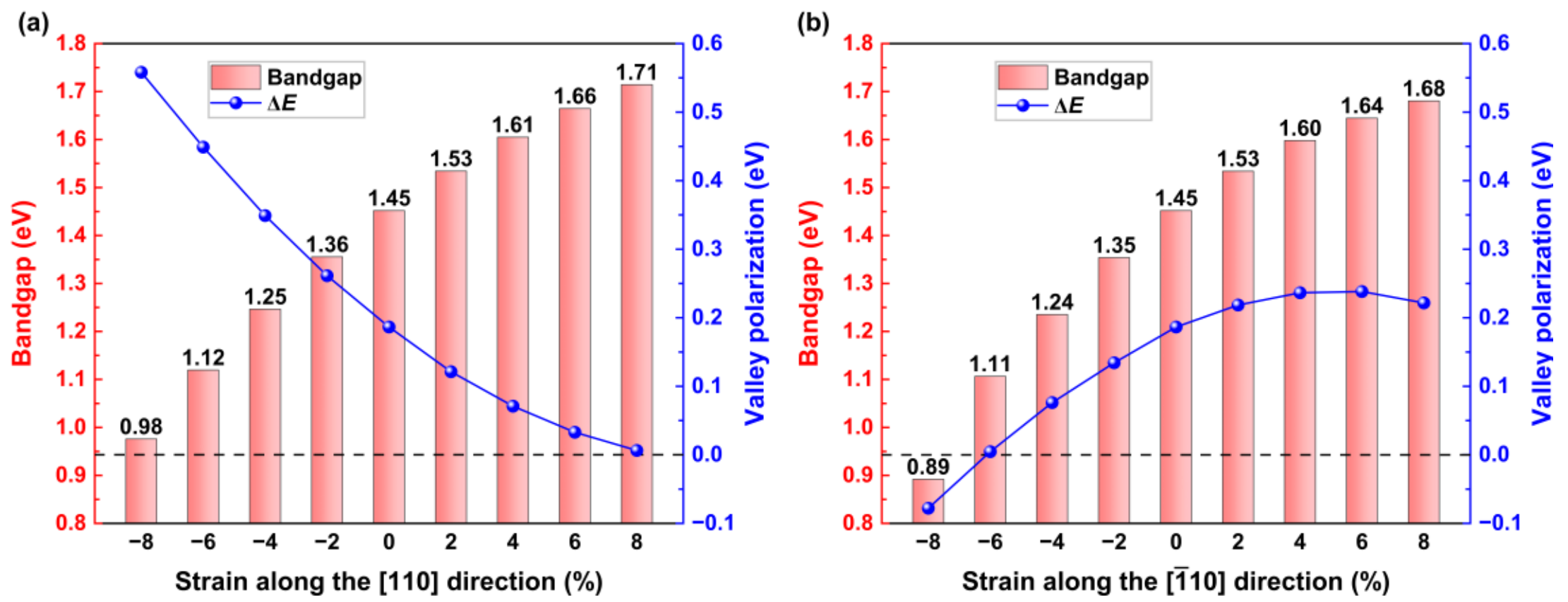

**FIG. 4.** (Color online) The bandgaps and the valley polarizations $\Delta E$ under strain conditions along the (a) [110] and (b) $[\bar{1}10]$ directions at the HSE06 level. The left axes represent the bandgap values, while the right axes indicate the values of $\Delta E$.

When compressive strain is applied along both directions, the bandgap decreases, and the energy of the Se-locked valley $E_{\mathrm{CB(Se)}}^{\mathrm{DFT}}$ and the Te-locked valley $E_{\mathrm{CB(Te)}}^{\mathrm{DFT}}$ both decreased. However, the effects are different: when compressive strain is applied along the [110] direction, the valley polarization $\Delta E$ increases, meaning that $E_{\mathrm{CB(Se)}}^{\mathrm{DFT}}$ decreases more rapidly. In contrast, when compressive strain is applied along the $[\bar{1}10]$ direction, the valley polarization $\Delta E$ decreases, meaning that $E_{\mathrm{CB(Te)}}^{\mathrm{DFT}}$decreases more rapidly. In contrast, along the $[\bar{1}10]$ direction, the perturbation diminishes and $\Delta E$ decreases, which means that $E_{\mathrm{CB(Se)}}^{\mathrm{DFT}}$ decreases more rapidly. From equation (1), one can find that the energy of valley state with a faster decrease will be the first to be restricted, and the corresponding BEA will be deactivated. The BEAs associated with each valley under strain conditions, compared with the redox potential of water at $p$H = 7 is shown in Fig. 5. In the ground state, both BEAs are activated. Under -4% strain along the [110] direction, only BEA 2 is activated, and the HER occurs on the upper surface. Under -2% strain along the [$\bar{1}$10] direction, only BEA 1 is activated, and the HER occurs on the lower surface. Both BEAs are deactivated in other cases. Therefore, a selective switch for the BEAs based on the applied in-plane uniaxial strain has been successfully realized, which allows for precise control of the HER site at the atomic scale, facilitating the spontaneous separation for the products of photocatalytic water splitting.

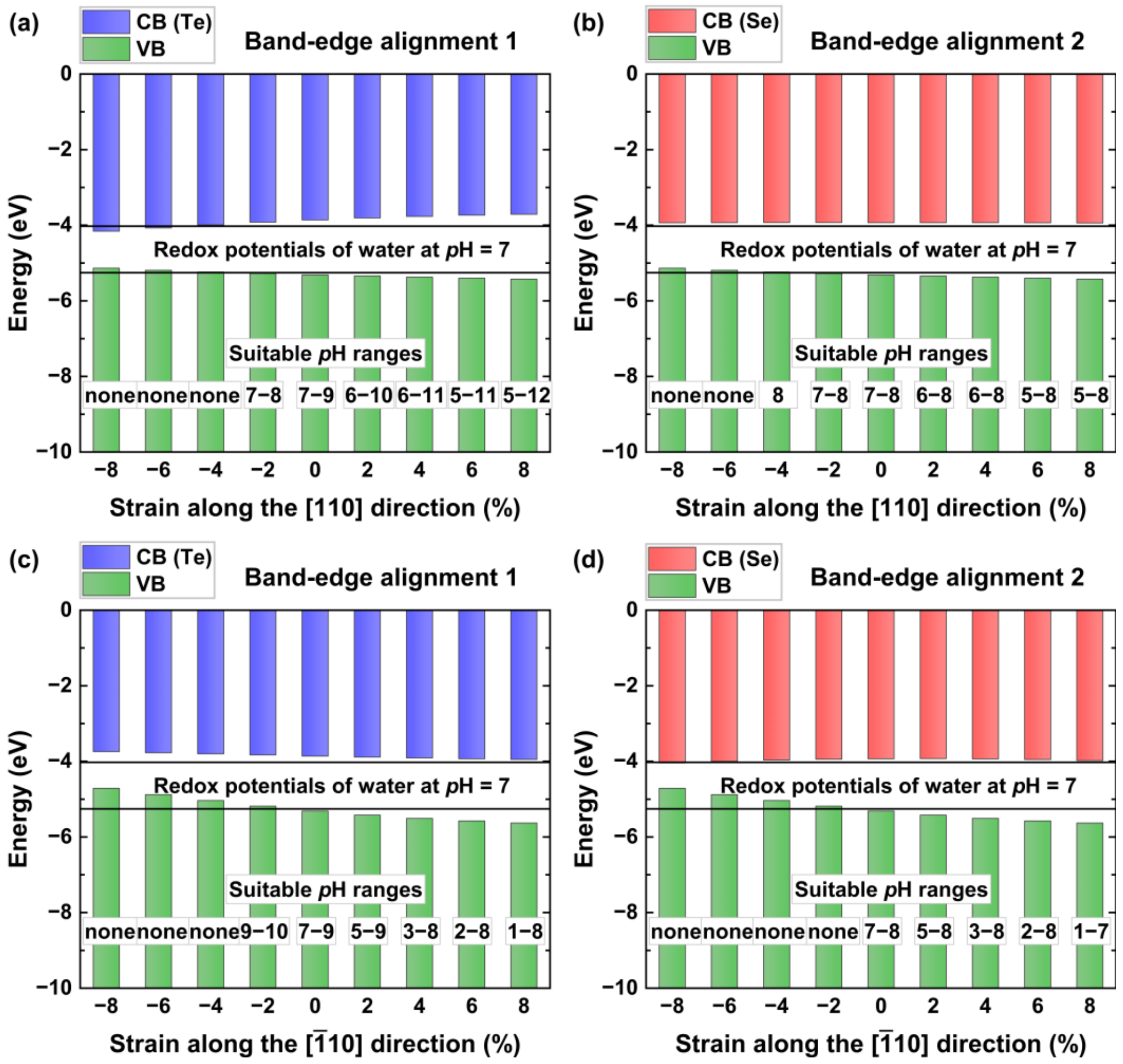

**FIG. 5.** (Color online) The band-edge alignment associated with each valley under strain conditions along the (a, b) [110] and (c,d) [$\bar{1}$10] directions, compared to the redox potential of water at *p*H = 7. The annotations indicate the suitable *p*H ranges for photocatalytic water splitting under strain conditions.

However, for a phenomenon that is only observed in monolayers, its application potential is limited due to the difficulty in synthesis. To address this, we designed four high-symmetry stacked bilayers based on the $CdSe_2$ monolayer, namely AA, AB, AA', and AB', with the AA stacking configuration having the lowest energy. The crystal structures, symmetries, and atom-resolved projected band structures of these configurations are shown in Fig. S15 in the Supplemental Material [36]. Under the high-symmetry AA and AB stackings, the structure retains S4 symmetry, thus satisfying the two requirements discussed earlier. In the AA stacking, the S4 axis passes through the Cd atoms, while in the AB stacking, the S4 axis passes through the midpoint of the Se-Se bond. AVL phenomena can be observed in the corresponding band structures. In contrast, for the AA' and AB' high-symmetry stackings, the absence of S4 symmetry results in valley polarization. However, these configurations cannot be considered perturbations of any structure that satisfies both requirements. In these cases, the atom-resolved projected band structures resemble those of conventional band structures, with no observable AVL phenomenon. These results naturally extend the concept of AVL to bilayer systems. Based on the discussion above, we predict that the AVL phenomena can be observed in any even-layered system that satisfies the symmetry requirement.

In conclusion, we proposed a novel concept of AVL and outlined its symmetry requirements. Through first-principles calculations, 72 candidate materials were screened. Using $CdSe_2$ monolayer as an example, we provide a detailed explanation of the unique role played by S4 symmetry in generating this atom-dependent valley distribution. Next, we designed CdSeTe monolayer to introduce an intrinsic EF into the system without destroy the AVL feature. A spontaneous valley polarization that does not involves time-reversal symmetry have been observed and valley-selected and face-specific HER site have been identified. In CdSeTe monolayer, two sets of valley electrons can be distinguished, each of them associated with a distinct BEA. Then, we designed a selective switch for BEAs through strain engineering, which enables precise atomic-scale manipulation of HER site. Finally, we extended the AVL concept to bilayer systems, enhancing the experimental feasibility of the proposed applications. These results indicate that studying novel valley behaviors in non-magnetic systems is beneficial and suggest that valleytronics hold great potentials in the field of photocatalytic water splitting.

**Acknowledgments**

This work was supported by the National Natural Science Foundation of China (Grant No. 12074274) and the Natural Science Foundation of Chongqing City (Grant No. cstc2020jcyjmsxmX0616).

**Conflict of interest**

There are no conflicts of interest to declare.

**Data availability**

The data that support the findings of this study are available from the corresponding author upon reasonable request.